\documentstyle[12pt,epsfig]{article}

\def\slashchar#1{\setbox0=\hbox{$#1$}           
   \dimen0=\wd0                                 
   \setbox1=\hbox{/} \dimen1=\wd1               
   \ifdim\dimen0>\dimen1                        
      \rlap{\hbox to \dimen0{\hfil/\hfil}}      
      #1                                        
   \else                                        
      \rlap{\hbox to \dimen1{\hfil$#1$\hfil}}   
      /                                         
   \fi}                                         %

\newcommand{\bc}{\begin{center}}
\newcommand{\ec}{\end{center}}
\newcommand{\be}{\begin{equation}}
\newcommand{\ee}{\end{equation}}
\newcommand{\bea}{\begin{eqnarray}}
\newcommand{\eea}{\end{eqnarray}}
\newcommand{\ba}{\begin{eqnarray}}
\newcommand{\ea}{\end{eqnarray}}
\newcommand{\brr}{\begin{array}}
\newcommand{\err}{\end{array}}

\newcommand{\simge}{\ \lower-
1.2pt\vbox{\hbox{\rlap{$>$}\lower5pt
\vbox{\hbox{$\sim$}}}}\ }

\begin{document}
\pagestyle{empty}
\vspace{-0.6in}
\begin{flushright}
BUHEP-01-31\\
CPT-2001/PE.4294\\
\end{flushright}
\vskip 0.8in
\centerline{\large {\bf{A Study of the 't Hooft Model}}}
\centerline{\large {\bf{with the Overlap Dirac Operator}}}
\vskip 0.6cm
\centerline{\bf{F.~Berruto$^{(\mbox{a})}$,
L.~Giusti$^{(\mbox{b})}$
, C.~Hoelbling$^{(\mbox{c})}$,
C.~Rebbi$^{(\mbox{a})}$}}
\vskip 1cm
\centerline{$^{(\mbox{a})}$Boston University - Department of Physics}
\centerline{590 Commonwealth Avenue, Boston MA 02215, USA}
\vskip 0.3cm
\centerline{$^{(\mbox{b})}$ 
Centre de Physique Theorique, CNRS Luminy, Case 907}
\centerline{F-13288 Marseille Cedex 9, France}
\vskip 0.3cm
\centerline{$^{(\mbox{c})}$John von Neumann Institute of Computing (NIC),
DESY Zeuthen}
\centerline{Platanenallee 6 D-15738 Zeuthen, Germany}
 \vskip 0.3cm
\centerline{e-mail:     fberruto@bu.edu}
\centerline{$\;\;\;\;\;\;\;\;\;\;\;\;\;\;\;\;\;\;\;\;\;\;\;$
Leonardo.Giusti@cern.ch}
\centerline{$\;\;\;\;\;\;\;\;\;\;\;\;\;\;\;\;\;\;\;\;\;\;\;\;\;\;\;$
Christian.Hoelbling@desy.de}
\centerline{$\;\;\;\;\;\;$ rebbi@bu.edu}
\vskip 0.6in
\begin{abstract}
\noindent
We present the results of an exploratory numerical study of 
two dimensional $QCD$ with overlap fermions.
We have performed extensive simulations for $U(N_c)$ and 
$SU(N_c)$ color groups with $N_c=2, 3, 4$ and 
coupling constants chosen to satisfy the 't Hooft 
condition $g^2 N_c =\mbox{const}=4/3$. 
We have computed the meson spectrum and decay constants, 
the topological susceptibility and the chiral condensate.
For $U(N_c)$ gauge groups, our results indicate that the 
Witten-Veneziano relation is satisfied within our statistical 
errors and that the chiral condensate for $N_f=1$ is compatible
with a non-zero value. Our results exhibit universality in~$N_c$ 
and confirm once more the excellent chiral properties of the
overlap-Dirac operator.
\end{abstract}
\vfill
\pagestyle{empty}\clearpage
\setcounter{page}{1}
\pagestyle{plain}
\newpage
\pagestyle{plain} \setcounter{page}{1}

\newpage

\section{Introduction}
\label{intro}
Several years ago, in a pioneering investigation, 't Hooft studied $U(N_c)$
gauge theories in the limit $N_c\rightarrow \infty$ with $C_t=g^2N_c$ kept
constant~\cite{thooft0}. He showed that only planar diagrams with quarks at
the edges dominate and therefore some non-perturbative $QCD$ physical 
observables can be computed in this limit. 
He proposed two dimensional models \cite{thooft1} 
with important features of QCD, but simple enough to sum explicitly all planar 
diagrams in the meson spectrum computation.

Recently, the overlap formulation~\cite{neuberger,neubnara} 
has made it possible to introduce a lattice Dirac operator $D$
which preserves a lattice form of chiral symmetry at
finite cut-off~\cite{luscher}. As a consequence the $U_A(1)$ chiral anomaly 
is recovered \`a la Fujikawa~\cite{F} and the Dirac Operator has exact chiral
zero modes for topologically non-trivial background configurations~\cite{laliena}. 
A precise and unambiguous implementation of the Witten--Veneziano formula 
can be obtained \cite{giusti}.

These theoretical developments generated a renewed
interest in 't Hooft's results and prompted us to perform an exploratory numerical
investigation of a class of two dimensional non-Abelian models with overlap
lattice fermions. Precisely, we have simulated models of $QCD_2$ with $U(N_c)$
and $SU(N_c)$ color groups for $N_c=2,3,4$ ($N_f=0,1,2$) imposing 
the 't Hooft's condition $C_t=g^2N_c=\mbox{const}$. Our systems are small enough 
that we could compute the fermionic propagator $D^{-1}$ and $\det(D)$ exactly following 
the scheme used in~\cite{rebbi1} (see Refs.~\cite{rebbi1,2o} and 
\cite{Edwards:1998yw,Hernandez:2001yn,Noi_bello,Dong:2001fm} for more refined 
implementations in two and four dimensions).

For $U(N_c)$ models we have found many background gauge configurations with
zero modes in the fermionic operator. By counting them and 
averaging over the configurations we have computed the quenched topological 
susceptibility obtaining values in very good agreement with the 
analytic results. We have computed the chiral condensate for $N_f=1$ 
which turns out to be compatible with a non-zero value.
For $SU(N_c)$ models we have not found any configuration with exact zero modes
as expected since these models have an exact $U_A(1)$ symmetry in the
chiral limit.

From two-point correlation functions of fermion bilinears 
we have extracted the meson masses and the corresponding decay
constants. In the $U(N_c)$ case the $\eta '$ mass in the chiral limit verifies 
the Witten-Veneziano relation~\cite{witten0,veneziano} within 
errors for each $N_c$. The pion masses verify quite well the 
expected functional dependence $M_{\pi}^2\propto m_q$~\cite{thooft1}. 
For $SU(N_c)$ models our data favor the functional dependence $M_{\pi}\propto m_q^{2/3}$ 
~\cite{steinhardt,hamer}. In both cases, at fixed $N_f$, data exhibit 
universality in~$N_c$ and quenched results get closer and closer to 
unquenched ones when $N_c$ increases.

In the next section we shall briefly remind the reader of some
properties of the 't Hooft model in the continuum. In section~\ref{odos} 
we define the overlap regularization we have implemented numerically. 
In section \ref{numres} we present our numerical
results for the meson spectra and decay constants.
In Section \ref{topsuwv} we compute the topological
susceptibility and we compare the $\eta'$ mass with the one 
extracted from the Witten-Veneziano formula.
In section \ref{chico} we report our results for
the chiral condensates. Section \ref{conclusions} is
devoted to some concluding remarks.

\section{The 't Hooft Model in the Continuum}
\label{overlap}
We consider two-dimensional models with
color group $U(N_c)$ and $SU(N_c)$ with $N_f$ degenerate
flavors defined by the action
\begin{equation}
S=\int d^2x \left[\frac{1}{2}\mbox{tr}F_{\mu\nu}F_{\mu\nu}
+\sum_{i=1,N_f}\overline{\psi}_i(\gamma_{\mu} D_{\mu}+m_q)
\psi_i\right]
\label{tha}
\end{equation}
The $\psi$ is a $N_f$-dimensional fermion multiplet\footnote{Color and spinor 
indices are suppressed throughout the paper.} and 
we use the following representation of the two-dimensional
$\gamma$-matrices
\begin{equation}
\gamma_1=\sigma_1\ ,\quad \gamma_2=\sigma_2\ ,\quad
\gamma_5=-i\gamma_1\gamma_2=\sigma_3
\label{gammas}
\end{equation}
where $\sigma_i$ are the Pauli matrices. In Eq.~(\ref{tha})
$D_{\mu}=\partial_{\mu}+igA_{\mu}$, where $A_{\mu}=A_{\mu}^A t^A$ is 
the gauge potential and $t^A$ are the $N_c^2$
or $N_c^2-1$ generators for the groups $U(N_c)$ or $SU(N_c)$
respectively, normalized according to $\mbox{tr}(t^A t^B)=1/2\ \delta^{AB}$. 
The field strength reads $F_{\mu \nu}=F^A_{\mu\nu}t^A$, where
\begin{equation}
F_{\mu\nu}^A=\partial_{\mu}A_{\nu}^A-\partial_{\nu}A_{\mu}^A-
g f^{ABC}A_{\mu}^B A_{\nu}^C
\label{fst}
\end{equation} 
The models described in Eq.~(\ref{tha}) are
super-renormalizable and therefore $g$ and $m_q$ are finite bare parameters.
't Hooft studied the $U(N_c)$ models in the limit 
\begin{equation}
N_c\rightarrow \infty\ ,\quad  g^2N_c=C_t=\mbox{constant}
\label{tholimit}
\end{equation}
which corresponds to take only planar diagrams with no fermion loops~\cite{thooft1}.

\subsection{The $U(N_c)$ models}
\label{uncm}
The massless action in Eq.~(\ref{tha}) has an $U_V(N_f)\otimes U_A(N_f)$ 
flavor symmetry. The $U_V(1)$ symmetry is preserved 
while the $U_A(1)$ is softly broken by the quark mass $m_q$
and explicitly broken by the anomaly which in two dimensions appears 
in two-point functions, not in triangle loops~\cite{thooft1}.
The corresponding singlet Ward identities are     
\begin{eqnarray}
\partial_{\mu}V_{\mu}(x)&=&0 \label{vwti}\\
\partial_{\mu}A_{\mu}(x)&=&2N_fQ(x)+2m_q P(x) \label{anomaly}
\end{eqnarray}
where
\begin{equation}
V_{\mu}(x)=\overline{\psi}(x)\gamma_{\mu}\psi(x)\ ,\quad 
A_{\mu}(x)=-i\epsilon_{\mu \nu}V_{\nu}(x)=\overline{\psi}(x)\gamma_{\mu}\gamma_5
\psi(x) \ ,\quad P(x)=\overline{\psi}(x)\gamma_5 \psi(x)
\label{vapc}
\end{equation}
and the topological charge density reads   
\begin{equation}
Q(x)=\frac{g\sqrt{N_c}}{4\pi}\ \epsilon_{\mu\nu}F_{\mu\nu}^0(x)
\label{topchargex}
\end{equation}
where $F_{\mu \nu}^0$ is the Abelian field strength. 
The Ward identities associated to the non-singlet ($N_f>1$) 
axial and vector symmetries are given by
\begin{eqnarray}
\partial_{\mu}V_{\mu}^f(x)&=&0\label{2vwi}\\
\partial_{\mu}A_{\mu}^f(x)&=&2m_q P^f(x) \label{2awi}
\end{eqnarray}
where $V_{\mu}^f$, $A_{\mu}^f$ and $P_{\mu}^f$ are defined
as in Eq.(\ref{vapc}), but with the insertion of the proper  
flavor generator. 

The topological charge of a $U(N_c)$ background gauge configuration is
\begin{equation}
Q=\int d^2 x\  Q(x)=
\frac{g\sqrt{N_c}}{4\pi}\int d^2x\ \epsilon_{\mu\nu}F_{\mu\nu}^0(x)
\label{topcharge}
\end{equation}
and it is related to the difference between the number of positive ($n_+$) and
negative ($n_-$) eigenvalues of the Dirac operator through the Atiyah-Singer
theorem
\begin{equation}
Q=n_--n_+
\label{astheo}
\end{equation}
In two dimensions the vanishing theorem ensures that
only $n_+$ or $n_-$ is non-zero~\cite{kiskis}.
The topological susceptibility $\chi$ in the pure gauge theory 
is~\cite{klebanov}
\begin{equation}
\chi =\int d^2x\ \langle
Q(x)Q(0)\rangle |_{YM}=\frac{C_t}{4\pi^2}
\label{topsu}
\end{equation}
where the expectation values $\langle \ldots \rangle |_{YM}$ 
in Eq.~(\ref{topsu}) have been taken in the theory without fermion fields. 

Since the $U_A(1)$ symmetry is anomalous, for $N_f=1$  one can have a 
non-zero chiral condensate~\cite{thooft1,berruto} 
$\langle \overline{\psi}\psi\rangle$ without 
violating the Mermin-Wagner theorem~\cite{coleman}. It is related to the  
topological susceptibility in the full theory as
\begin{equation}
\chi =-m_q\langle \bar{\psi}\psi \rangle
+O(m_q^2)
\label{topcond}
\end{equation}
On the contrary, for $N_f>1$ a non-zero condensate would spontaneously 
break the $SU_A(N_f)$ symmetry.

The meson spectrum of the two-flavor $U(N_c)$ models exhibits 
a pseudoscalar flavor-singlet excitation ($\eta'$) and flavor-triplet 
quark-antiquark bound states (pions). The $\eta'$ is massive due 
to the anomaly~\cite{thooft1} and its mass in the chiral limit is 
\begin{equation}
M_{\eta '}^2=N_f\ \frac{g^2}{\pi}=N_f\ \frac{C_t}{N_c}\ \frac{1}{\pi}
\label{ametanf}
\end{equation}
The Witten-Veneziano
formula~\cite{witten0,veneziano} for the $U(N_c)$ models gives
\begin{equation}
M_{\eta '}^2=\frac{4N_f}{f_{\pi}^2}\chi
\label{wvr}
\end{equation}
where $f_{\pi}$ is the pion decay constant. 
By inverting Eq.(\ref{wvr}), assuming that it is an exact relation between the
$\eta'$ mass and the topological susceptibility and
taking into account Eqs.~(\ref{topsu}), (\ref{ametanf}) and (\ref{wvr})
one gets
\begin{equation}
f_{\pi}=\sqrt{\frac{N_c}{\pi}}
\label{fpi}
\end{equation}
In the $N_c\rightarrow \infty$ limit
and for $m_q\ll \sqrt{C_t/\pi}$, the pion mass $M_{\pi}$ as a function of 
the quark mass $m_q$ reads~\cite{thooft1}
\begin{equation}
M_{\pi}^2=2\ \sqrt{\frac{C_t\  \pi}{3}}\ m_q+\ldots
\label{mpuvsmq}
\end{equation}
It is interesting to note that $M_{\pi}^2$ is linear at the leading order in 
the quark mass like in four-dimensional $QCD$ and the coefficient in front 
of $m_q$ is expected to be exact in the $N_c\rightarrow \infty$ 
limit~\cite{thooft1}.  

\subsection{The $SU(N_c)$ models}
\label{suncm}
Analogously to the previous case, 
the massless action (\ref{tha}) of two dimensional $SU(N_c)$ models
has an exact $U_V(N_f)\otimes U_A(N_f)$ flavor symmetry. Since 
in this case there is no Abelian field strength component $F_{\mu \nu}^0$,
the $U_A(1)$ symmetry is only softly broken by the quark mass $m_q$.
Therefore the Dirac operator should not have any zero-modes.

In the limit $N_c\rightarrow \infty$, 
a non-zero chiral condensate was obtained in~\cite{zhitnitsky}  
\begin{equation}
\langle \overline{\psi}\psi
\rangle=-N_c(\frac{g^{2}N_c}{12\pi})^{\frac{1}{2}}=-\frac{g}{\sqrt{12\pi}}N_c^{\frac{3}{2}}
\label{zchi}
\end{equation}
and a Berezinski-Kosterlitz-Thouless phase transition was
advocated to reconcile this result with the
Mermin-Wagner theorem~\cite{coleman}. 
This behavior would favor the argument that in the
limit $N_c\rightarrow \infty$ both $U(N_c)$ and $SU(N_c)$ gauge groups should
describe the same physics~\cite{thooft0,witten2,coleman2}.

The meson spectrum of the $SU(N_c)$ models exhibits only pions.
For $N_f=1$ and the $SU(2)$ color group 
their mass has been computed in ~\cite{hamer} in the semiclassical 
WKB approximation  
\begin{equation}
M_{\pi}^2=\frac{9}{\pi}\ (2^7C_t)^{\frac{1}{3}}
(\frac{e^{\gamma}}{\pi})^{\frac{4}{3}}\ m_q^{\frac{4}{3}}+\ldots
\label{mpsvsmq}
\end{equation}

\section{The Overlap Dirac operator}
\label{odos}
We have implemented the lattice action
\begin{equation}
S=S_G(U)+\sum_{i=1,N_f}\sum_{x,y}\overline{\psi}_i(x)D_{m_q}(x,y)\psi_i(y)
\label{la}
\end{equation}
where $S_G(U)$ is the standard Wilson gauge action
\begin{equation}
S_G(U)=\beta\sum_{x,\mu<\nu}\left[1-\frac{1}{2N_c}\mbox{Tr}\ ( U_{\mu
\nu}(x) + U_{\mu\nu}^\dagger(x) )\right]
\label{gwa}
\end{equation}
$\beta=2N_c/(ga)^2$, $a$ and $g$ being the lattice spacing and bare
coupling constant and
\begin{equation}
U_{\mu\nu}(x)=U_{\mu}(x)U_{\nu}(x+a\hat{\mu})U_{\mu}^{\dagger}(x+a\hat{\nu})
U_{\nu}^{\dagger}(x)
\label{plaquette}
\end{equation}
In Eq.~(\ref{la}) 
\begin{equation}
D_{m_q}(x,y)=\left[ (1-\frac{m_qa}{2})D(x,y)+m_q\ \delta_{xy}\right]
\label{mneuop}
\end{equation}
where the Neuberger-Dirac operator is defined as~\cite{neuberger} 
\begin{equation}
D \equiv \frac{1}{a}\left(1+(D_W-\frac{1}{a})\left[(D_W^{\dagger}-\frac{1}{a})
(D_W-\frac{1}{a})  \right]^{-\frac{1}{2}}\right)
\label{neuop}
\end{equation}
$D_W$ is the Wilson-Dirac operator   
\begin{equation}
D_W=\frac{1}{2}\gamma_{\mu}(\nabla_{\mu}+\nabla_{\mu}^*)-\frac{1}{2}a
\nabla_{\mu}^*\nabla_{\mu}
\label{wdo}
\end{equation}
where
\begin{eqnarray}
\nabla_{\mu}\psi_i(x)&=&\frac{1}{a}\left[U_{\mu}(x)\psi_i(x+a\hat{\mu})-\psi_i(x)
\right]\\
\nabla_{\mu}^*\psi_i(x)&=&\frac{1}{a}\left[\psi_i(x)-U_{\mu}^{\dagger}
(x-a\hat{\mu})\psi_i(x-a\hat{\mu})\right]
\end{eqnarray}
The Neuberger Dirac operator is $\gamma_5$-hermitian, i.e. $D^{\dagger}
=\gamma_5 D \gamma_5$, and satisfies the Ginsparg-Wilson
relation~\cite{ginsparg}
\begin{equation}
\gamma_5 D^{-1}+D^{-1}\gamma_5=a\gamma_5
\label{gwr}
\end{equation}
which guarantees that in the chiral limit the lattice action (\ref{la})
is invariant under the continuum symmetry~\cite{luscher}
\begin{equation}
\delta \psi_i=\gamma_5(1-aD)\psi_i,\quad \delta
\overline{\psi_i}=\overline{\psi_i}\gamma_5
\label{lusy}
\end{equation}
As a consequence the $U_A(1)$ anomaly, if present, 
is recovered \`a la Fujikawa~\cite{F} and the Dirac operator has exact chiral
zero modes for topologically non-trivial gauge field configurations~\cite{laliena}. 
The analogous flavor non-singlet chiral transformations are obtained by 
including a flavor group generator in Eq.~(\ref{lusy}). 

From Eq.~(\ref{gwr}) one can derive the following identities ($a=1$)
\begin{equation}
D^{\dagger}_{m_q}D_{m_q}=(1-\frac{m_q^2}{4})\ \left[
D+D^{\dagger}\right]+m_q^2
\label{ddagdm}
\end{equation}
\begin{equation}
D^{-1}_{m_q}=\left(\frac{1}{1+\frac{m_q}{2}}D^{\dagger}+\frac{m_q}
{1-\frac{m_q^2}{4}}\right)\frac{1}{\left[D+D^{\dagger}\right]
+m_q^2\left(1-\frac{m_q^2}{4}\right)^{-1}}
\label{invmneuop}
\end{equation}
which turns out to be useful in the numerical implementation.

\section{The Pion Masses and Decay Constants}
\label{numres}
In order to study the response of the overlap Dirac operator in the
presence and absence of the chiral anomaly and 
at the same time to analyze the scaling of the physical observables with $N_c$
(at fixed $N_f$), we have performed extensive simulations of  
$U(N_c)$ and $SU(N_c)$ models for $N_c=2, 3, 4$. 
We have generated the gauge configurations 
with a standard Metropolis Monte Carlo algorithm
according to the gauge action in Eq.~(\ref{gwa}).
To avoid the Gross-Witten phase transition~\cite{grosswitten}
we have chosen $C_t=g^2N_c=4/3$, which corresponds to $\beta=6$ for $N_c=2$,
$\beta=13.5$ for $N_c=3$ and $\beta=24$ for $N_c=4$. 
We have generated $500$ independent configurations for $N_c=2$,
$300$ for $N_c=3$ and
$150$ for $N_c=4$  separated by  $10000$ sweeps of the 
whole lattice. For all gauge groups we have fixed the same 
dimensions in lattice units, i.e. $N_t=N_x=18$. 
Consequently the Neuberger-Dirac operator is a
complex matrix of dimension $1296\times1296$ for $N_c=2$, $1944\times 1944$
for $N_c=3$ and $2592\times 2592$ for $N_c=4$.  We could diagonalize
exactly the hermitian operator $D+D^{\dagger}$
by using full matrix algebra routines with the resources available
to us. By using Eq.~(\ref{ddagdm}) we have computed 
the eigenvalues of the massive Neuberger operator and its determinant
and with Eq.~(\ref{invmneuop}) we have determined the propagators 
of the massive fermions. We have explicitly checked that, for
each of the fermionic masses $m_q=0.04,0.05,0.06,0.07,0.08,0.1$,
the lattice spans at least four pion correlation lengths.
The effects of dynamical fermions have been included by weighting the
observables with the appropriate powers of the fermion determinant. The
smallness of the $18\times 18$ lattice warrants this
procedure~\cite{rebbi1}, which would lead to an unacceptable large
variance on larger systems. 

\begin{figure}[!ht]
\begin{center}
\includegraphics[height=8cm,width=10cm]{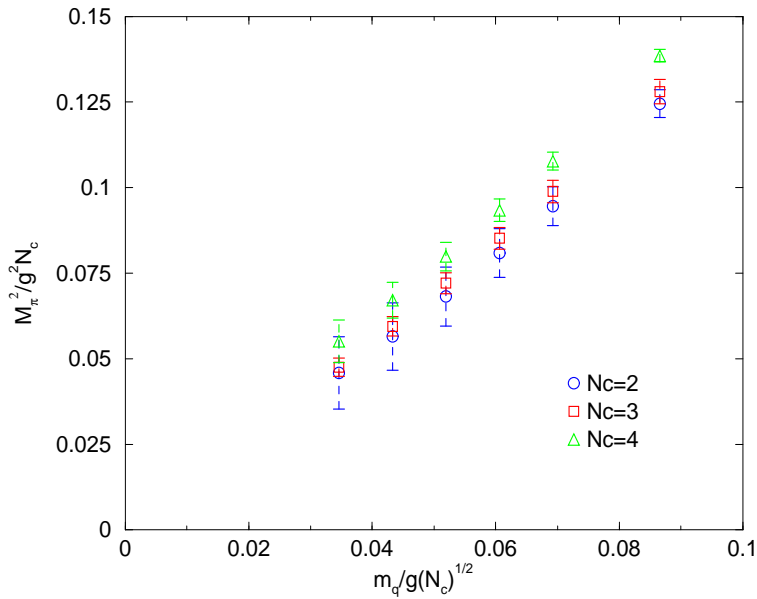}
\caption{\label{figpnf2u}
\it{$M^2_\pi/g^2N_c$ vs.~$m_q/g\sqrt{N_c}$ for $N_f=2$, $U(N_c)$ models.}}
\end{center}
\end{figure}

The meson masses and the decay constants have been extracted in the standard manner
from the vector correlator at zero momentum
\begin{equation}
\sum_{x,x',y} \langle  V_1(x,y) V_1(x',y+t) \rangle
\label{mvc}
\end{equation}
where the non-singlet vector current we have used is 
\begin{equation}
V_\mu(x,y) = \bar\psi_1(x,y) \gamma_\mu[(1-\frac{a}{2}D)\psi_2](x,y)
\end{equation}
\label{pions}
\begin{table}[!ht]
\begin{center}
\begin{tabular}{||l|l|l|l||}
\hline\hline
                 & $N_c=2$          & $N_c=3$          & $N_c=4$          \\
$m_q/g\sqrt{N_c}$&$M_{\pi}^2/g^2N_c$&$M_{\pi}^2/g^2N_c$&$M_{\pi}^2/g^2N_c$\\
\hline
\multicolumn{4}{||c||}{$N_f=2$}  \\
\hline
0.0346 & 0.046(10) & 0.047(3) & 0.055(6) \\
0.0433 & 0.056(10)  & 0.059(3) & 0.067(5) \\
0.0519 & 0.068(8)  & 0.072(3) & 0.080(4) \\
0.0606 & 0.080(7)  & 0.085(3) & 0.093(3) \\
0.0692 & 0.094(6)  & 0.098(3) & 0.107(3) \\
0.0866 & 0.124(4)  & 0.128(3) & 0.138(2) \\
\hline
\multicolumn{4}{||c||}{quenched ($N_f=0$)} \\
\hline
0.0346 &   0.069(1)   & 0.063(1)  &  0.057(1)   \\
0.0433 &   0.083(1)   & 0.077(1)  &  0.071(1)   \\
0.0519 &   0.098(1)   & 0.091(1)  &  0.086(1)   \\
0.0606 &   0.113(1)   & 0.106(1)  &  0.100(1)   \\
0.0692 &   0.129(1)   & 0.121(1)  &  0.116(1)   \\
0.0866 &   0.161(1)   & 0.153(1)  &  0.148(1)   \\
\hline\hline
\end{tabular}
\caption{\label{mp}
$M_{\pi}^2/g^2N_c$ vs. $m_q/g\sqrt{N_c}$, $U(N_c)$ models.}
\end{center}
\end{table}
In Table \ref{mp} we report the pion mass squared for
the $U(N_c)$ models with two flavors of dynamical fermions and
in the quenched approximation. The quenched results get closer 
to the unquenched two-flavor ones when $N_c$ gets larger.
Figure~\ref{figpnf2u} shows $M_{\pi}^2$ as a function of the quark mass
for $N_f=2$ and provides evidence of universality in 
$N_c$ for the pion masses.
According to Eq.~(\ref{mpuvsmq}), $M_{\pi}^2$   
should have a linear dependence in $m_q$ and it is 
expected to vanish in the chiral limit.
From a fit 
\begin{equation}
\frac{M_{\pi}^2}{g^2N_c}=A+B\ \frac{m_q}{g\sqrt{N_c}}
\label{ufitp}
\end{equation}
we obtained  $A=-0.013(17)$ and $B=1.57(19)$
for $N_c=2$, $A=-0.007(4)$ and
$B=1.54(4)$ for $N_c=3$ and $A=-0.006(8)$ and 
$B=1.66(9)$ for $N_c=4$. 
In order to carefully determine the systematic error affecting $B$, further
simulations would be necessary that go beyond the scope of the present 
investigation.

The pion decay constants $f_{\pi}$ turns
out to be almost constant in $m_q$ within our statistical errors.
Their values in the chiral limit are reported in Table~\ref{fpitab}. 
\begin{table}[!ht]
\begin{center}
\begin{tabular}{||l|l|l||}
\hline\hline
$N_c=2$          & $N_c=3$          & $N_c=4$\\
$\quad f_{\pi}$ & $\quad f_{\pi}$ & $\quad f_{\pi}$       \\
\hline
\multicolumn{3}{||c||}{$N_f=2$}    \\
\hline
0.7(1) &     0.94(5) &   1.10(1) \\
\hline
\multicolumn{3}{||c||}{quenched ($N_f=0$)}     \\
\hline
0.77(1)  &     0.96(1)  &   1.12(1)  \\
\hline\hline
\end{tabular}
\caption{\label{fpitab} $f_{\pi}$, $U(N_c)$ models.}
\end{center}
\end{table}

\begin{figure}[!ht]
\begin{center}
\includegraphics[height=8cm,width=10cm]{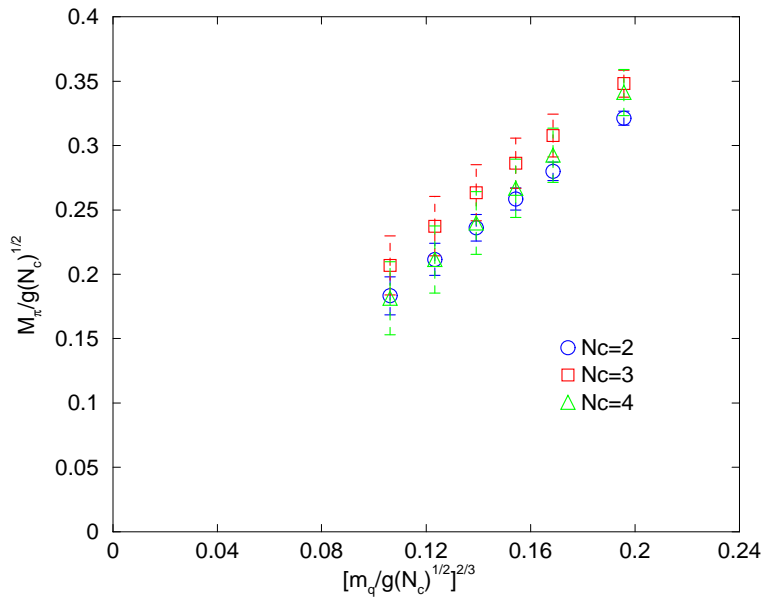}
\caption{\label{figpnf2s}
\it{$M_\pi/g\sqrt{N_c}$ vs.~$(m_q/g\sqrt{N_c})^{2/3}$ for $N_f=2$,
$SU(N_c)$ models.}}
\end{center}
\end{figure}

In Table~\ref{mps} we report the pion
masses for the $SU(N_c)$ models with one and two flavor of dynamical fermions
and in the quenched approximation.
\begin{table}[!ht]
\begin{center}
\begin{tabular}{||l|l|l|l||}
\hline\hline
                 & $N_c=2$          & $N_c=3$          & $N_c=4$          \\
$(m_q/g\sqrt{N_c})^{2/3}$&$M_{\pi}/g\sqrt{N_c}$&$M_{\pi}/g\sqrt{N_c}$&$M_{\pi}/g\sqrt{N_c}$\\
\hline
\multicolumn{4}{||c||}{$N_f=2$} \\
\hline
0.1063  &  0.183(15)  &  0.207(23)   & 0.181(28)    \\
0.1233  &  0.211(12)  &  0.237(23)   & 0.212(26)    \\
0.1392  &  0.236(10)  &  0.263(22)   & 0.240(24)    \\
0.1543  &  0.259(8)   &  0.286(19)   & 0.267(23)    \\
0.1687  &  0.280(7)   &  0.308(16)   & 0.292(21)    \\
0.1957  &  0.321(5)   &  0.348(10)   & 0.341(18)    \\
\hline
\multicolumn{4}{||c||}{$N_f=1$} \\
\hline
0.1063  &  0.182(7)  & 0.196(4)  &  0.199(12) \\
0.1233  &  0.211(6)  & 0.226(4)  &  0.229(10) \\
0.1392  &  0.237(5)  & 0.254(4)  &  0.258(9)  \\
0.1543  &  0.262(4)  & 0.279(3)  &  0.285(8)  \\
0.1687  &  0.285(4)  & 0.303(3)  &  0.310(7)  \\
0.1957  &  0.329(3)  & 0.349(2)  &  0.357(6)  \\
\hline
\multicolumn{4}{||c||}{quenched ($N_f=0$)} \\
\hline
0.1063  & 0.201(3)  & 0.212(1)   & 0.216(1)  \\
0.1233  & 0.227(2)  & 0.241(1)   & 0.246(1)  \\
0.1392  & 0.253(2)  & 0.268(1)   & 0.273(1)  \\
0.1543  & 0.277(2)  & 0.293(1)   & 0.299(1)  \\
0.1687  & 0.300(2)  & 0.316(1)   & 0.323(1)  \\
0.1957  & 0.344(2)  & 0.361(1)   & 0.368(1)  \\
\hline\hline
\end{tabular}
\caption{\label{mps}
$M_{\pi}/g\sqrt{N_c}$ vs. $(m_q/g\sqrt{N_c})^{2/3}$, $SU(N_c)$ models.}
\end{center}
\end{table}
Figure~\ref{figpnf2s} shows the pion masses for $N_f=2$ and reveals also in this 
case a universality in $N_c$ already for $N_c=2,3,4$. 
According to Eq.~(\ref{mpsvsmq}), $M_{\pi}^2$  
should have an $m_q^{2/3}$ quark mass dependence and should vanish
for $m_q=0$. From a fit
\begin{equation}
\frac{M_{\pi}}{g\sqrt{N_c}}=A+B\ (\frac{m_q}{g\sqrt{N_c}})^{\frac{2}{3}}
\label{sfitp}
\end{equation}
we got $A=0.023(26)$ and $B=1.52(12)$
for $N_c=2$, $A=0.046(43)$ and
$B=1.55(17)$ for $N_c=3$ and $A=-0.01(4)$ and 
$B=1.79(12)$ for
$N_c=4$, where again the errors are statistical only. 
We performed also a fit of the form
\begin{equation}
M_{\pi}/g\sqrt{N_c}=C\left(m/g\sqrt{N_c}\right)^{\gamma}
\end{equation}
and we obtained a value of $\gamma$ compatible with $2/3$ for $N_c=2,3,4$.

The values of the pion decay constants $f_{\pi}$ 
in the chiral limit for the $SU(N_c)$ models are reported in Table~\ref{fpitab_SU}. 
\begin{table}[!ht]
\begin{center}
\begin{tabular}{||l|l|l||}
\hline\hline
$N_c=2$          & $N_c=3$          & $N_c=4$\\
$\quad f_{\pi}$ & $\quad f_{\pi}$ & $\quad f_{\pi}$       \\
\hline
\multicolumn{3}{||c||}{$N_f=2$}    \\
\hline
 0.64(6) &  0.92(3) &  1.21(12)\\
\hline
\multicolumn{3}{||c||}{$N_f=1$}     \\
\hline
 0.70(2)&  0.98(2)&  1.14(3)\\
\hline
\multicolumn{3}{||c||}{quenched ($N_f=0$)}     \\
\hline
 0.76(1)  &  0.98(1) &    1.12(1)\\
\hline\hline
\end{tabular}
\caption{\label{fpitab_SU} $f_{\pi}$, $SU(N_c)$ models.}
\end{center}
\end{table}

\section{The Witten-Veneziano Relation and the $\eta '$ Mass}
\label{topsuwv}
The topological charge of a given background configuration can be 
computed by counting the number of zero modes of the 
overlap Dirac operator. Since we have diagonalized 
exactly the overlap operator, we could compute the quenched topological 
susceptibility $\chi$ by averaging the square 
of the topological charge, normalized with the volume, over the configurations. 
In Table~\ref{toposus} we 
report the results we have obtained for the various $N_c$.
\begin{table}[!ht]
\begin{center}
\begin{tabular}{||l|l|l||}
\hline\hline
$N_c=2$          & $N_c=3$          & $N_c=4$          \\
$\quad \chi$     & $\quad \chi$        & $\quad \chi$ \\
\hline
 & & \\
0.0258(16) & 0.0298(23) & 0.0319(37) \\
 & & \\
\hline\hline
\end{tabular}
\caption{\label{toposus}Topological susceptibility, $U(N_c)$ models.}
\end{center}
\end{table}

Even if with increasing $N_c$ the topological susceptibility get closer and closer to the
analytical value $\chi = 0.0337$ given in Eq.~(\ref{topsu}),
more accurate studies of discretization and finite size effects, 
that goes beyond the scope of our exploratory investigation, are required
to reach a final conclusion. In fact if the magnitude of these systematic 
uncertainties is different for the gauge groups, they would spoil 
any extrapolation to $N_c\rightarrow \infty$.
For example, building on a trade-off 
between spatial and internal
degrees of freedom that prompted the introduction of the $N_c\rightarrow
\infty$ single plaquette model~\cite{eguchi}, one could heuristically argue
that the theory should be less affected by finite size effects for increasing
$N_c$. To try to estimate finite size effects in the worst case, 
we computed the value of the topological susceptibility of
the $U(2)$ model for the volume $V=24\times 24$ and we obtained
$\chi =0.0316(24)$. The difference of this value
with the analogous one in Table~\ref{toposus} gives a rough estimate
of the systematic error induced by finite volume effects on $\chi$.

For $SU(N_c)$ model we directly checked that there are no zero modes 
for each gauge field configuration and therefore the topological 
charge is always zero.
\begin{table}[!ht]
\begin{center}
\begin{tabular}{||l|l|l||}
\hline\hline
$N_c=2$          & $N_c=3$          & $N_c=4$          \\
$M_{\eta '}^2/g^2N_c$ & $M_{\eta '}^2/g^2N_c$ & $M_{\eta '}^2/g^2N_c$ \\
\hline
\multicolumn{3}{||c||}{Analytic}\\
\hline
0.159   & 0.106   &0.079         \\
\hline
\multicolumn{3}{||c||}{Numerical from Eq.(\ref{wvr})}\\
\hline
0.129(6)&0.097(7) &0.076(6) \\
\hline
\multicolumn{3}{||c||}{Numerical from Eq.(\ref{etarebbi})}\\
\hline
 0.159(8)  & 0.111(7) & 0.084(8)     \\
\hline\hline
\end{tabular}
\caption{\label{wvrtab} $M_{\eta '}^2$ from the Witten-Veneziano relation
in Eq.~(\ref{wvr}) and from Eq.(\ref{etarebbi}).}
\end{center}
\end{table}

From the values of $\chi$ reported in Table~\ref{toposus} and the pion
decay constants at $N_f=0$ given in Table~\ref{fpitab}, we have 
computed $M_{\eta '}^2$ for $N_f=1$ given in Table~\ref{wvrtab} (statistical errors only). 
It turns out that the WV relation is satisfied for all 
$U(N_c)$ gauge groups within our overall uncertainties.    

The singlet pseudoscalar 
correlation functions are given by differences between connected and 
disconnected contributions and therefore they are noisier with respect 
to the non-singlet case. To have an independent determination of the
$\eta'$ mass for $N_f=1$, we resorted the method proposed in~\cite{hamber}. 
It exploits the quenched two-loop disconnected
$\Gamma_q^{\mbox{2-loop}}$ and one-loop connected
$\Gamma_q^{\mbox{1-loop}}$ contributions to the $\eta '$ propagator
\begin{equation}
\Gamma_q(t)=\int dx\ \langle P(x,t)P(0,0)\rangle\Big{|}_{\rm{quenched}}
\label{etaprop}
\end{equation}
where the singlet pseudoscalar density is defined as 
\begin{equation}
P(x,t) = \bar\psi(x,t) \gamma_5[(1-\frac{a}{2}D)\psi](x,t)
\end{equation}
The formula we have used reads ~\cite{hamber}
\begin{equation}
M_{\eta '}^2=2M_{\pi}\lim_{t\rightarrow\infty}
\frac{\Gamma_q^{\mbox{2-loop}}(t)}
{|t|\Gamma_q^{\mbox{1-loop}}(t)}
\label{etarebbi}
\end{equation}
The results we have obtained, reported in Tab.~\ref{wvrtab},
compares remarkably well with the analytic result in the chiral limit.

\section{Chiral condensate}
\label{chico}
The chiral condensate can be expressed in terms of the 
eigenvalues $\lambda_i=1+e^{i\theta_i}$
of the Neuberger-Dirac operator as ($a=1$)
\begin{figure}[!ht]
\begin{center}
\includegraphics[height=8cm,width=10cm]{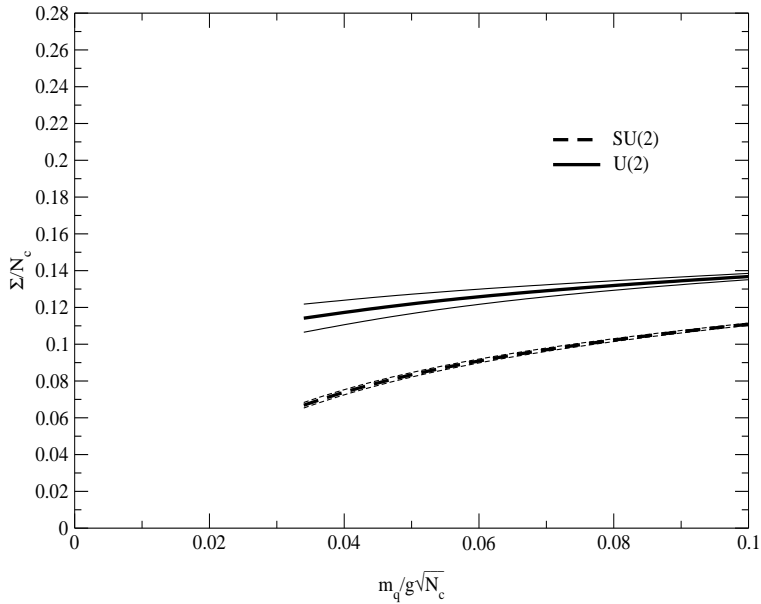}
\caption{\label{usu2}$N_f=1$ chiral condensate for $U(2)$ and $SU(2)$.}
\end{center}
\end{figure}
\begin{equation}
\Sigma=-\frac{1}{N_f g\sqrt{N_c}}\langle \bar{\psi}(1-\frac{1}{2}D)\psi \rangle
=\frac{1}{g\sqrt{N_c} V}
\frac{\langle (\mbox{det}
  D_{m_q})^{N_f} F(Q,m_q,\cos \theta_i) \rangle_U}
{\langle(\mbox{det} D_{m_q})^{N_f}\rangle_U}
\label{chicond}
\end{equation}
where
\begin{equation}
F(Q,m_q,\cos \theta_i)=\frac{|Q|}{m_q}+\frac{m_q}{2}\sum_{i, \cos
  \theta_i\not
= -1}
\frac{1-\cos \theta_i}{(1+\frac{m_q^2}{4})+(1-\frac{m_q^2}{4})\cos \theta_i}
\label{F}
\end{equation}
and using Eq.~(\ref{ddagdm}) 
\begin{equation}
(\mbox{det}
D_{m_q})^{N_f}=m_q^{N_f|Q|}
\prod_{i,\ \cos\theta_i\not= -1}
\left[\left(1-\frac{m_q^2}{4}\right)2(1+\cos \theta_i)+m_q^2
\right]^{\frac{N_f}{2}}
\label{detfer}
\end{equation}
For all the groups we have computed 
the chiral condensate by using the direct 
definition in Eq.~(\ref{chicond}).
As a representative example,
in Fig.~\ref{usu2} we compare the
values of the $U(2)$ and $SU(2)$ chiral condensates for $N_f=1$ computed up to 
$m_q/g\sqrt{N_c}=0.0346$ in order to limit the finite size effects. 
A chiral extrapolation of these plots hints to a
finite and zero value of the $U(2)$ and $SU(2)$ chiral condensates,
respectively.

\begin{figure}[!ht]
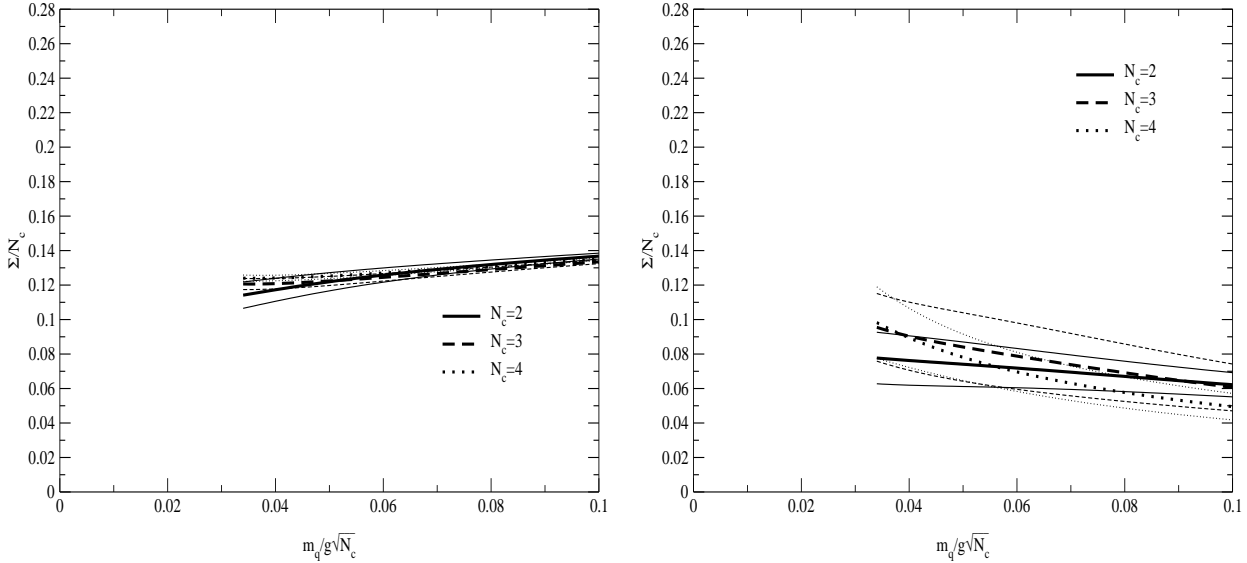

\begin{center}
\begin{tabular}{cc}
\mbox{\epsfig{file=fig_new_5a.eps,height=7.5cm,width=8.0cm,angle=0}} &
\mbox{\epsfig{file=fig_new_5b.eps,height=7.5cm,width=8.0cm,angle=0}} \\
\end{tabular}
\caption{$N_f=1$ chiral condensate. Left: from Eq.~(\ref{chicond}). 
Right: from Eq.~(\ref{topcond}).
\label{u234}}
\end{center}
\end{figure}

In Fig.~\ref{u234} we show the chiral condensate for 
all $U(N_c)$ models with $N_f=1$. They have been computed 
by using the definition in Eq.~(\ref{chicond}) and the one from the 
Axial Ward Identity in Eq.~(\ref{topcond}) obtained 
by neglecting the $O(m_q^2)$. The comparison is 
interesting because the mass dependence of the two quantities 
is different. Again both the determinations point in the direction 
of non-zero condensates.
An interesting improvement in the determination of the condensates 
in the chiral limit could be obtained by implementing in two dimensions
a finite volume technique analogous to the one proposed in 
four dimensions in Refs.~\cite{Leutwyler:1992yt,HJL}. Moreover the use of 
algorithms which generate directly dynamical configurations 
could help in reducing the fluctuations induced by the re-weighting of the 
observables with the fermionic determinant.

\section{Conclusions}
\label{conclusions}
We have performed an exploratory numerical study on the lattice of two dimensional 
models defined by the gauge groups $U(N_c)$ and $SU(N_c)$  ($N_c=2,3,4$) 
and $N_f=0,1,2$ degenerate fermions introduced by using the Neuberger-Dirac 
operator. Our results prove that the computation is feasible and it would be
interesting to further pursue this line of research with a more detailed
analysis, especially of discretization and finite size effects, 
that goes beyond the scope of our present investigation.

We have found that within our statistical errors the pion masses verify quite well the 
expected functional dependence  $M_{\pi}^2\propto m_q$ and 
$M_{\pi}\propto m_q^{2/3}$ for $U(N_c)$ and $SU(N_c)$ models respectively. 
In both cases, at fixed $N_f$, data exhibit 
universality in~$N_c$ and quenched results get closer and closer to 
unquenched ones when $N_c$ increases.

As expected from an analysis of the symmetries of the models, 
for $SU(N_c)$ groups we have not found any background gauge configuration with
exact zero modes in the fermionic operator. On the other hand, 
many background gauge configurations with exact zero modes
were found for the $U(N_c)$ models. By counting them and 
averaging over the configurations we have computed the quenched topological 
susceptibility obtaining values in good agreement with the 
analytic results. By using the meson decay constants extracted from the 
two-point functions, our data verify the Witten-Veneziano 
relation within errors for each $N_c$.
Even if we could not safely extrapolate our data to the chiral limit,
we have produced evidences that the chiral condensate for $N_f=1$ 
is compatible with a non-zero value.

Of course, one would be interested to know how close real $QCD$ is to the
large $N_c$ limit. In Refs.~\cite{teper} it was pointed out, on the
basis of a study of three and four dimensional pure gauge theories,
that even the $SU(2)$
color group is close to $SU(\infty)$. Although our analysis is limited to
$QCD_2$, remarkably enough we saw that even with dynamical quarks
2, 3 and 4 colors appear to be in the same universality class, 
i.e. the physical quantities are degenerate in $N_c$ within the errors.

Although our study is exploratory, the results we have obtained 
are very gratifying and indicate that it would be interesting to
perform a deeper analysis in two dimensions and eventually  
extend it to four dimensions. 
In particular one could study within our computational scheme the baryons
that are expected to be the $QCD$ solitons~\cite{witten2}. 
An interesting improvement in the determination of the condensates 
in the chiral limit could be obtained by implementing 
finite volume techniques in two dimensions. Moreover the use of 
algorithms which generate  directly dynamical configurations
could help in reducing the fluctuations induced by the re-weighting of the 
observables with the fermionic determinant and therefore would allow 
to simulate larger lattices.

\section*{Acknowledgments}
We warmly thank G.C.~Rossi for many illuminating discussions.
We wish to thank A.~Gonzalez-Arroyo and H.~Neuberger for
interesting discussions. This work has been supported in part under DOE grant
DE-FG02-91ER40676. The work of F.B.~is supported in part by an INFN
Postdoctoral Fellowship. He thanks the members of the Theoretical Particle
Physics Group at Boston University for their kind hospitality.

\end{document}